\documentclass[12pt, draftclsnofoot, onecolumn]{IEEEtran}

\usepackage{cite}
\usepackage{graphicx}
\usepackage{amsfonts}
\usepackage{amsmath}
\usepackage{amssymb}
\usepackage{amsbsy}
\usepackage{bm}
\usepackage{multicol}
\usepackage{multirow}
\usepackage{psfrag}
\usepackage{stfloats}
\usepackage{algorithmic}
\usepackage{algorithm}
\usepackage{array}
\usepackage{color}
\usepackage{booktabs}

\graphicspath{{figure/}}

\begin{document}
	
\title{Physical Layer Security in UAV Systems: Challenges and Opportunities}

\author{Xiaofang Sun,~\IEEEmembership{Member,~IEEE,} 
	Derrick Wing Kwan Ng,~\IEEEmembership{Senior Member,~IEEE,}\\
	Zhiguo Ding,~\IEEEmembership{Senior Member,~IEEE,}
	Yanqing Xu,~\IEEEmembership{Student Member,~IEEE,}\\
	and~Zhangdui Zhong,~\IEEEmembership{Senior Member,~IEEE}
	\thanks{X. Sun, Y. Xu, and Z. Zhong are with the State Key Lab of Rail Traffic Control and Safety and the Beijing Engineering Research Center of High-speed Railway Broadband Mobile Communications, Beijing Jiaotong University, Beijing 100044, China (emails: \{xiaofangsun, yanqing\_xu, zhdzhong\}@bjtu.edu.cn).}	
	\thanks{D. W. K. Ng is with the School of Electrical Engineering and Telecommunications, University of New South Wales, Sydney, NSW 2052, Australia (Email: w.k.ng@nusw.edu.au).}	
	\thanks{Z. Ding is with the School of Electrical and Electronic Engineering, The University of Manchester, Manchester, UK (Email: zhiguo.ding@manchester.ac.uk).}
}

\maketitle
\begin{abstract}
	
	Unmanned aerial vehicle (UAV) wireless communications have experienced an upsurge of interest in both military and civilian applications, due to its high mobility, low cost, on-demand deployment, and inherent line-of-sight (LoS) air-to-ground channels.
	However, these benefits also make UAV wireless communication systems vulnerable to malicious eavesdropping attacks.
	In this article, we aim to examine the physical layer security issues in UAV systems.
	In particular, passive and active eavesdroppings are two primary attacks in UAV systems.
	We provide an overview on emerging techniques, such as trajectory design, resource allocation, and cooperative UAVs, to fight against both types of eavesdroppings in UAV wireless communication systems.
	Moreover, the applications of non-orthogonal multiple access, multiple-input and multiple-output, and millimeter wave in UAV systems are also proposed to improve the system spectral efficiency and to guarantee security simultaneously.
	Finally, we discuss some potential research directions and challenges in terms of physical layer security in UAV systems.
\end{abstract}

\section{Introduction}

In the past few years, heterogeneous data applications have emerged in wireless communications, such as video streaming, e-health monitoring, video conferencing, etc.
However, traditional communication networks with fixed infrastructures are unable to meet the increasingly stringent quality-of-service (QoS) requirements in the fifth-generation (5G) and beyond 5G networks.
As a result, the development of unmanned aerial vehicle (UAV) has created a fundamental paradigm shift in wireless communication systems to facilitate fast and highly flexible deployment of communication infrastructures.
In particular, by exploiting the high maneuverability of UAV, communication links can be established ubiquitously, especially in temporary hotspots, disaster areas, and complex terrains.
Compared to traditional terrestrial wireless communications, UAV wireless communications have the following unique features \cite{Zeng_UAV_2016}:
\begin{itemize}
	\item \textbf{On-demand and swift deployment:} UAV enables fast establishment of temporal communication infrastructures in emergency scenarios where legacy or fixed infrastructures are destroyed or do not exist, such as disaster rescue, remote sensing, firefighting, and others.  The deployment of UAV facilitates cost-efficient and uninterrupted communications.
	\item \textbf{High flexibility:} Due to the fully controllable three-dimensional (3D) mobility, UAV can either stay quasi-stationarily or cruise continuously to a dedicated location, depending on the requirements of wireless communication systems. The movement of UAVs provides a new degree of freedom to offer efficient communications. 	
	\item \textbf{High probability of line-of-sight (LoS) air-to-ground  channel:} According to the field trial results \cite{3GPP}, the LoS component dominates the air-to-ground channels in many practical scenarios, especially for rural areas or moderately height UAV altitude. 
	This channel characteristic leads to the fact that the channel state information (CSI) can be directly determined by the position of each node and used to facilitate the design of high-speed communication systems.
	\item \textbf{Limited resources:} It is noted that onboard batteries of UAVs have limited energy storage. Moreover, the sizes of UAVs are usually small for on-demand deployment, and the weights of UAVs for communications are typically not exceeding $25$ kg for both safety and energy consumption issues \cite{Zeng_UAV_2016}. These limitations directly restrict the communication, computation, and cruising duration capabilities of UAV.
\end{itemize}

Despite the promising gains brought by UAVs, the open nature of air-to-ground wireless channels makes secure information transfer a challenging issue.
Specifically, on the one hand, information signals transmitted over wireless LoS channels are likely to be intercepted by some undesired receivers which leads to a risk of information leakage. On the other hand, wireless UAV transceivers are vulnerable to malicious jamming attacks.
Hence, security plays an extremely important role in UAV wireless communications. Unfortunately, traditional encryption techniques require high computational complexity leading to a large amount of energy consumption \cite{Survey_PLS_2014} which may not be suitable to UAV systems.
As an alternative, physical layer security is computationally efficient and effective in safeguarding wireless communication networks via exploiting the inherent randomness of wireless channels.
As a result, various physical layer techniques have been proposed in the literature for guaranteeing communication security, e.g., \cite{Survey_PLS_2014,Cai_JSAC_2018,Li_2018,Xiaofang_GC,Worst_Case,UAV_Jamming_TVT_2018,PLS_Cooperative,Massive_MIMO_ActiveEve,Jonas_CM_2015}.

Since the channel characteristics determine the performance of physical layer security, the LoS channels as well as the mobility and flexibility of UAVs bring both opportunities and challenges into the physical layer security design in UAV systems.
In particular, on the one hand, the LoS channel condition in UAV-based communication systems may increase the vulnerability to eavesdropping.
On the other hand, the fully controllable mobility of UAVs can be exploited to enhance communication security via adjusting its trajectory.
For example, if the locations of eavesdroppers are known, the UAV can avoid flying close to them to reduce the potential of information leakage for guaranteeing secure transmission.
Hence, introducing UAV to wireless  communication systems is a double-edged sword which requires a careful system design and thorough research.

Motivated and inspired by the unique challenges and opportunities brought by UAV, 
this paper aims to provide an important overview on physical layer security in UAV wireless communication systems, which sheds light on potential researches in this field.
In particular, the potential security attacks specifically in UAV wireless communication systems are discussed. 
	Moreover, corresponding solutions are provided in terms of emerging techniques by considering both the advantages and limitations of UAV. 
	To further enhance the physical layer security, the applications of advanced 5G technologies, e.g., NOMA, 3D beamforming, and mmWave, are investigated in UAV wireless communication systems. 
	Finally, potential research challenges that have not been addressed in the literature and some future directions in terms of secure communications in UAV systems are envisioned.

\section{Security Attacks in UAV Systems}\label{Sec:Secure_Attacks}

A typical physical layer security communication problem comprises a minimum of three nodes, i.e., a legitimate transmitter, a legitimate receiver, and a potential eavesdropper, which can be modeled by a wiretap channel \cite{Survey_PLS_2014}.
In general, secrecy capacity and secrecy outage probability are the two fundamental metrics to evaluate the physical layer security performance.
Typically, denote $C_{\mathrm{M}}$ and $C_{\mathrm{E}}$ as the Shannon capacities of the main channel and eavesdropping channel, respectively. Consequently, the secrecy capacity $C_{\mathrm{S}}$ is defined by \cite{Survey_PLS_2014}
\begin{align}\label{eq:Secrecy_Capacity}
C_{\mathrm{S}} = [C_{\mathrm{M}}-C_{\mathrm{E}}]^+,
\end{align}
where $[x]^+\triangleq \max\{x,0\}$.
Notably, perfectly secure communication between a legitimate pair is possible, when the eavesdropper's channel is a degraded version of the main channel.

\begin{figure}
	\begin{center}
		\includegraphics[width=0.95\columnwidth]{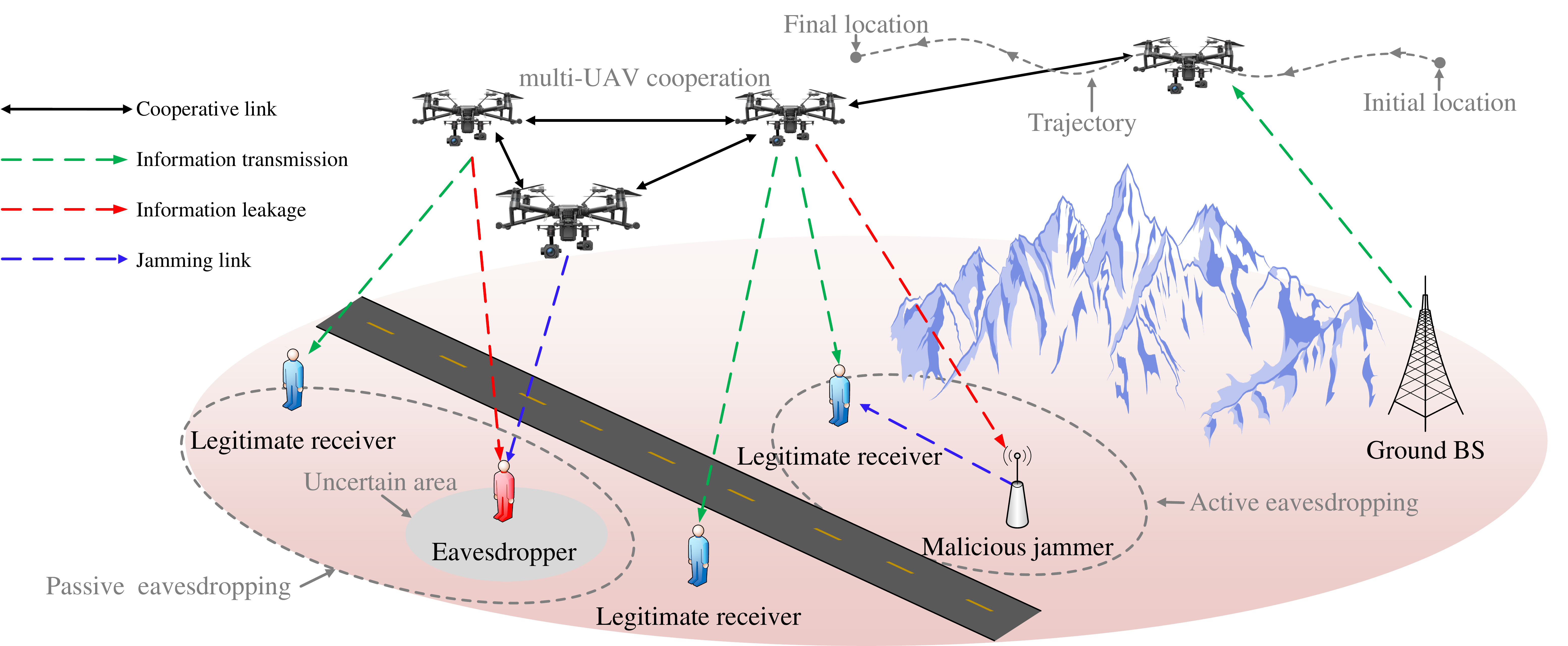}
		\caption{Illustrations of security issues in UAV wireless communication systems.}\label{fig:UAV_PLS}
	\end{center}
\end{figure}

In UAV-based communication systems, a UAV can act as either a transmitter or a receiver.
On the one hand, it is noted that when the UAV is a legitimate transmitter, the associated air-to-ground LoS channels facilitate the signal reception at both legitimate receiver and the eavesdropper, which may increase the vulnerability to potential eavesdropping. 
On the other hand, when the UAV acts as a legitimate receiver, the existence of LoS channels enhance the physical layer security especially against passive eavesdropping.
As such, to investigate the security issues in the presence of eavesdroppers, we mainly focus on the former scenario where the UAV acts as a legitimate transmitter.
The potential security attacks in UAV systems are illustrated in Figure \ref{fig:UAV_PLS} and will be described in the following sections.

\subsection{Passive Eavesdropping Scenarios}

In this scenario,  passive eavesdroppers intend to intercept some confidential data without degrading the received signal quality at the legitimate receiver.
Since the eavesdroppers work in a passive manner and only intend to intercept the confidential messages, they usually remain silent and  their position information is not easy to be obtained by the UAV, which makes the system  vulnerable to eavesdropping.
In the following, we discuss three cases with different availabilities of the eavesdroppers' position information.

\subsubsection{Full Position Information of Eavesdroppers}

An optical camera or a synthetic aperture radar equipped at the UAV can help to detect and track the positions of potential  external eavesdroppers.
If the eavesdroppers stay stationary, obtaining the complete positions of these external eavesdroppers are possible at the UAV.
As such,  the security issues are likely to be addressed in such scenarios via proper resource allocation design and taking advantage of the flexibility of UAV.
However, obtaining such precise position information requires a high hardware cost.
Besides, the required new equipments impose an extra load to UAV which may increase its energy consumption.


\subsubsection{Partial Position Information of Eavesdroppers}

	When the eavesdroppers stay quasi-stationary or move in certain areas, obtaining precise position information of eavesdroppers is challenging and expensive in terms of hardware cost and energy consumption.
	Then, with the aid of the equipped camera or radar on the UAV, partial position information of these eavesdroppers is possible to be obtained at the UAV via sensing and tracking.
	In the literature, worst-case secrecy capacity and secrecy outage probability are the two common metrics to evaluate the performance of UAV systems with partial position information of external eavesdroppers \cite{Survey_PLS_2014,Worst_Case,UAV_Jamming_TVT_2018}.

\subsubsection{Absence of Position Information of Eavesdroppers}

	When eavesdroppers hide themselves physically, it is difficult for the UAV to detect and track them. 
	This raises a problem that the completely absence of position information makes the UAV system extremely vulnerable to eavesdropping. 	
	In such scenarios, perfectly secure communications cannot be always guaranteed.

\subsection{Active Eavesdropping Scenarios}

Compared to passive eavesdroppings, active ones are more dangerous.
The reasons are twofold.
On the one hand, the active eavesdropping  intends to attack the main channel by degrading the channel capacity.
For instance, active eavesdroppers transmit jamming signals to the legitimate  receiver for degrading the capacity of the main channel, i.e., $C_{\mathrm{M}}$.
On the other hand, the active eavesdropping may also aim to improve the capacity of the eavesdropping channel.
As such, eavesdropping attacks caused by an active eavesdropper is more harmful than that of passive ones.
Specifically, the LoS channel characteristic can improve the capacity of the eavesdropping channel when the UAV is a legitimate transmitter.
According to the duplexing mode, we discuss the following two types of jamming attacks in UAV wireless communication systems:
\begin{itemize}
	\item {\it Full-duplex (FD) eavesdropper:}
	A FD active eavesdropper transmits jamming  noise and intercepts confidential signals simultaneously and independently \cite{Chenxi_TWC}.
	In such scenarios, the UAV would increase its transmit power to improve the quality of the jammed channel which not only facilitates the eavesdropping at the FD eavesdropper due to the LoS channel characteristics, but also consumes more energy to combat the active eavesdroppings.
	
	\item {\it Cooperative half-duplex (HD) eavesdroppers:}
	Multiple HD eavesdroppers can mimic a FD eavesdropper via cooperation.
	For instance, some eavesdroppers transmit jamming to the legitimate receiver, while the others intercept the confidential signal.
\end{itemize}

In both situations, when the CSI of the legitimate receiver is available at the active eavesdroppers, the UAV systems become severely vulnerable, as the active eavesdroppers can efficiently interfere the legitimate receiver.

\section{Physical Layer Security Techniques for  UAV Systems}

To enhance the physical layer security, the unique properties of UAVs, e.g., high mobility and flexibility in positioning, can be exploited.
Moreover, by using advanced resource allocation techniques, the secrecy performance can be further improved.
In the following, we discuss the specific effective physical layer security strategies against the corresponding eavesdroppers listed in Section \ref{Sec:Secure_Attacks} in air-to-ground channels, as shown in Table \ref{Tab}.

\begin{table}[!t]
	\centering
	\caption{Specific Security Attacks and Possible Solutions in UAV Wireless Communication Systems}\label{Tab}
	\begin{tabular}{|c|c|c|c|c|c|c|c|}
		\hline
		\multicolumn{2}{|c|}{\multirow{2}{*}{Security attacks}}                                                                               & \multicolumn{3}{c|}{Passive eavesdroppings}                                                             & \multicolumn{3}{c|}{FD/Cooperative HD active eavesdroppings}                                         \\ 
		\cline{3-8}   
		\multicolumn{2}{|c|}{}                                    
		& Full PI      & Partial PI    & Absence of PI  
		& Full PI    & Partial PI    & Absence of PI  \\
		\hline
		\multirow{4}{*}{Techniques}
		&Joint design
		&\checkmark & \checkmark & \checkmark  &\checkmark &\checkmark  &\checkmark \\ \cline{2-8} 
		&Robust design
		& &\checkmark& &  &\checkmark &  \\ \cline{2-8}
		&Artificial noise
		& & &\checkmark &\checkmark &\checkmark &\checkmark \\ \cline{2-8}
		&Multi-UAV CoMP &\checkmark &\checkmark &\checkmark &\checkmark &\checkmark &\checkmark\\ 
		\cline{1-8}		
		\multirow{3}{*}{Technologies} 
		&Multi-antenna
		  & \checkmark                       & \checkmark & \checkmark & \checkmark    &\checkmark & \checkmark                                 \\ 
		\cline{2-8} 
		& NOMA                    & \checkmark                       & \checkmark &                           &            & &                                                  \\ \cline{2-8} 
		& mmWave  & \checkmark    & \checkmark & \checkmark & \checkmark   & \checkmark & \checkmark                                  \\ \hline
		\multicolumn{8}{l}{\begin{tabular}[c]{@{}l@{}}
				PI: Position information\\
				Joint design: Joint trajectory and resource allocation design
		\end{tabular}}                                                          
	\end{tabular}
\end{table}

\subsection{Anti-Eavesdropping Techniques}

\subsubsection{Joint Trajectory and Resource Allocation Design}

In UAV systems, there are various resources needed to be allocated, such as transmit power, cruising speed, time slot, and frequency bandwidth. 	
It is noted that resource allocation affects the signal strength received at not only the legitimate receiver, but also the eavesdroppers.
In the following, we first discuss how trajectory design can be exploited to improve the physical layer security in UAV systems.
We then provide an overview on the joint design of resource allocation and trajectory for physical layer security provisioning.

\paragraph{Trajectory Design Approach}

In practice, the initial and final locations of the UAV are usually predetermined. Besides, the UAV is also constrained by the maximum cruising speed and duration.
These define the maximum service area.
As a result, UAV trajectory design has become an important research topic for realizing efficient UAV-based communication systems \cite{Cai_JSAC_2018,Li_2018,Xiaofang_GC}.
In terms of communication security, the UAV can fly close to the legitimate ground node and away from the eavesdropper if it is possible.
The principle is to  carefully design the trajectory of UAV, such that the legitimate link can be enhanced while the eavesdropping link becomes weaken.

\paragraph{Resource Allocation}

Specifically, a joint design of trajectory and resource allocation is a promising approach to further enhance physical layer security, e.g.,  \cite{Li_2018,Xiaofang_GC}.
The basic principle of the joint design is that when a UAV has to fly close to the eavesdropper, the UAV can decrease or shut down the transmission power to reduce the potential of information leakage.
At the same time, the UAV flies away from the eavesdropper with its full speed for saving more time slots for the future.
In contrast, when the UAV flies close to the legitimate receiver, the UAV usually slows down and increases its transmit power for the confidential information transmission.

We provide an example to show how UAV can be adopted as a mobile relay to guarantee secure communications via joint trajectory and resource allocation design.
In the considered scenario, a legitimate transmitter intends to serve a legitimate receiver in the presence of a passive eavesdropper with the aid of a UAV.
The UAV is introduced as a mobile relay to complete the data delivery and to enable physical layer security.
The ultimate goal of the system design is to maximize the spectral efficiency of the system, while guaranteeing secure communications and subject to  certain practical constraints as in \cite{Xiaofang_GC}.
The joint trajectory and resource allocation design can be obtained by iteratively solving the corresponding approximated optimization problem.

\begin{figure}[!t]
	\begin{minipage}{0.5\columnwidth}
		\centering
		\includegraphics[width=1\linewidth]{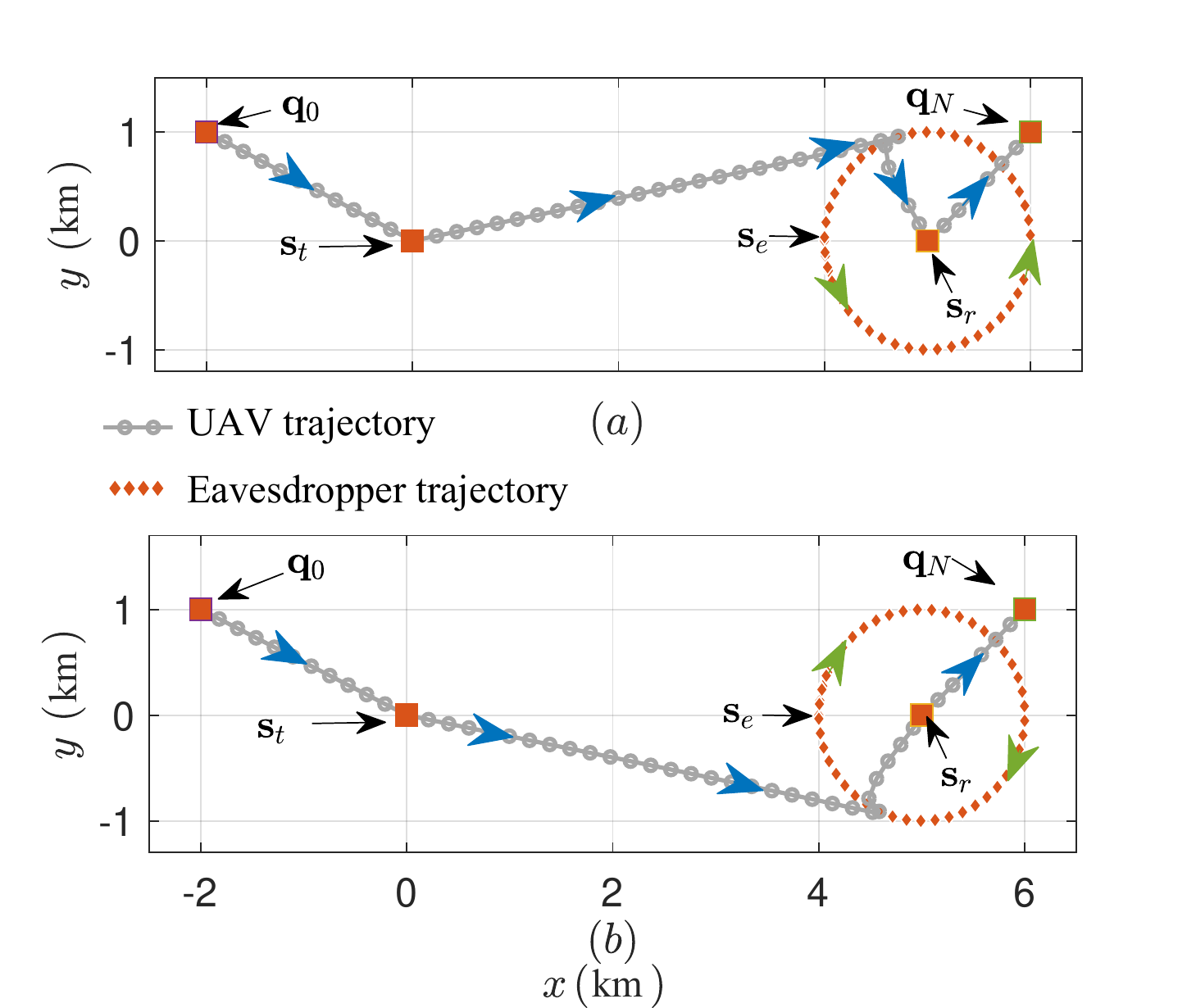}	
	\end{minipage}
	\begin{minipage}{0.5\columnwidth}
		\centering
		\includegraphics[width=1\linewidth]{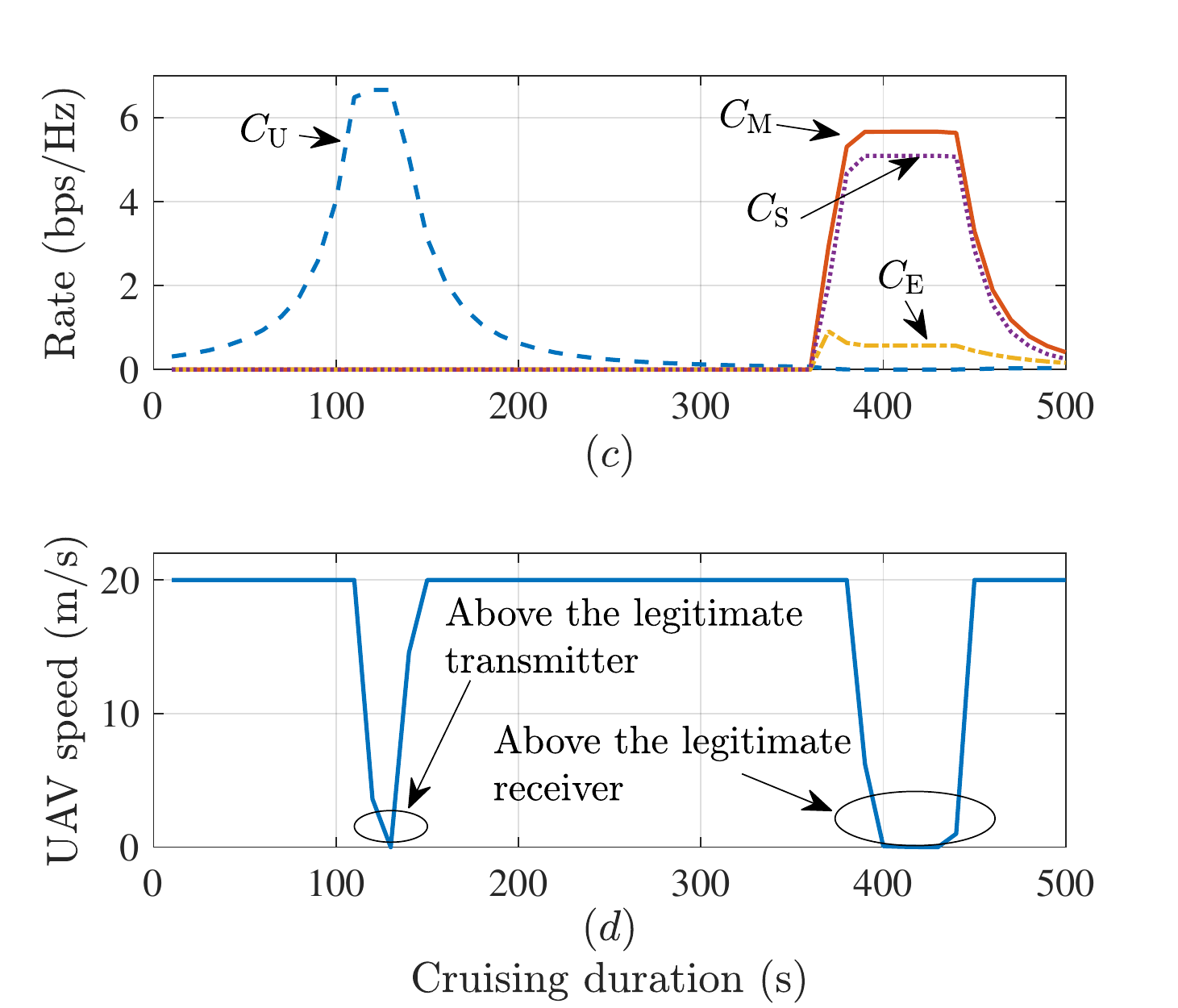}
	\end{minipage}
	\caption{The optimized UAV trajectories for the eavesdropper moves anticlockwise (a) and clockwise (b) around the legitimate receiver with radius 1 km.
		The achieved system rate (c) and the corresponding cruising speed (d) via joint trajectory and resource allocation design to enhance the physical layer security in UAV-aided wireless communication scenarios.}
	\label{fig:Joint_Design}
\end{figure}

Figure \ref{fig:Joint_Design} provides some interesting insights into the trajectory and transmission performance during the operation period.
In this figure, the legitimate transmitter and its receiver locate at $(0, 0, 0)$ and $(5$ km, $0,0)$, respectively, in a 3D Cartesian coordinate system.
The position of the eavesdropper is assumed to be available at the UAV for the optimal design. The maximum cruising speed of the UAV is $20$ km/h.

Figure \ref{fig:Joint_Design}(a) and Figure \ref{fig:Joint_Design}(b) depict the optimized trajectories of a UAV for two interesting scenarios, where the eavesdropper dynamically moves anticlockwise and clockwise, respectively, around the legitimate receiver with radius $1$ km and velocity $50$ km/h.
Variables $\mathbf{s}_t$, $\mathbf{s}_r$, and $\mathbf{s}_e$ represent locations of the legitimate transmitter,  legitimate receiver, and eavesdropper, respectively. $\mathbf{q}_0$ and $\mathbf{q}_N$ represent the predetermined initial and final locations of UAV, respectively.
From this figure, we find that the UAV first flies close to the transmitter for caching enough data.
Then it approaches the legitimate receiver and avoids any possible trajectories leading close to the eavesdropper.
Figure \ref{fig:Joint_Design}(c) and Figure \ref{fig:Joint_Design}(d) depict the capacities and the corresponding cruising speed of the UAV, respectively, via the optimal joint design.
In these two figures, the eavesdropper locates stationary at ($4$ km, $0, 0)$.
$C_\mathrm{U}$ denotes the instantaneous achievable rate at the UAV.
From Figure  \ref{fig:Joint_Design}(c) and Figure  \ref{fig:Joint_Design}(d), we find that when the UAV locates closer to the eavesdropper than to the legitimate receiver, the UAV either caches data from the legitimate transmitter by slowing down and then hovering above it or keeps silent and moves away from the eavesdropper.
Differently, when the UAV is closer to the legitimate receiver, the UAV transmits confidential signal with a positive secrecy rate and tries to hover above the desired receiver as long as possible.
These indicate that joint trajectory and resource allocation design in UAV systems is an effective way to enhance the physical layer security and improve spectral efficiency.

\subsubsection{Robust Joint Design}

When only partial or statistical position information of eavesdroppers is available, robust joint design can be exploited to facilitate secure UAV communications by considering the worst case scenario.
	For example, the uncertain location area of the potential eavesdropper can be modeled by a region with a center which is the exact location of the eavesdropper where the length of its radius is related to the amount of uncertainty. 
	Then, the robust design for the worst-case communication security is to guarantee the QoS of the system whenever the eavesdropper is located within the region.
	In general, the UAV with optimized trajectory would fly as close as possible to the legitimate receiver for enhancing the capacity of the legitimate channel while cruises away from the uncertain area of the eavesdropper(s) as much as possible. 
	Also, a higher transmit power from the UAV is adopted when the UAV is close to the legitimate user to exploit the short distance communication between the transceivers.
However,  when the UAV  has to approach the uncertain area of the eavesdropper(s), it decreases or shuts down the transmit power and accelerates away from this area to reduce the potential of information leakage.

\subsubsection{Artificial Noise}\label{Subsubsec:AN}

To address the security issues caused by passive eavesdroppers with the absence of their position information, one effective approach is to transmit artificial noise to the null space of the legitimate pair's channels via performing cooperation among multiple UAVs and to transmit confidential data only when UAVs are close enough to the legitimate receiver via joint trajectory and resource allocation design.
To this end, confidential data and artificial noise are available among the cooperating UAVs.
As such, the capacity of the confidential signal through the legitimate channel, i.e., $C_{\mathrm{M}}$ in Eq. \eqref{eq:Secrecy_Capacity}, is increased but that through the eavesdropping channel, i.e., $C_{\mathrm{E}}$, is reduced. However, since  artificial noise consumes some transmit power,  this would leave a smaller amount of system power for allocating to the confidential information signal. 
Therefore, power allocation is an important issue in UAV-based wireless communications for improving the system secrecy capacity. 
However, finding the optimal power allocation for UAV-based typically means solving non-trivial NP-hard optimization problems which are generally intractable. 
Hence, computationally efficient suboptimal solutions should be proposed.

\subsection{Anti-Jamming Techniques}

To address the security issues caused by jamming attacks, cooperation among multiple UAVs \cite{PLS_Cooperative} can be exploited to enhance the physical layer security, 
since when multiple UAVs are available, cf. Figure \ref{fig:UAV_PLS}, the degrees of freedom for optimizing the system resources are increased.
The cooperation approaches can be summarized into the following two cases.

For instance, cooperative multi-point (CoMP) transmission technique can be employed at UAVs.
In particular, these multiple UAVs can form a virtual antenna-array to enhance the received signal strength at legitimate receivers while degrade that at the eavesdroppers.
Furthermore, with multiple UAVs in the system, one can optimize their trajectories and resource allocations such that some UAVs transmit confidential signal to the ground legitimate receivers while the others send jamming signals to confuse the eavesdroppers.
According to the availabilities of eavesdroppers' positions, UAV can adopt different transmission schemes against active eavesdroppings:
\begin{itemize}
	\item When UAVs have the complete position information of the eavesdroppers, multiple UAVs can be employed to facilitate a joint trajectory and resource allocation design via forming a virtual antenna-array. 
	Consequently, the confidential signal transmission can be enhanced by focusing the information energy beams to the ground legitimate receiver while reducing the possibility of information leakage.
	\item When UAVs have only partial position or statistical information of the eavesdropper, robust joint trajectory and resource allocation design can be introduced into multi-UAV networks to improve the secrecy rate of the system or to reduce the secrecy outage probability below a certain level \cite{Worst_Case}.
	\item 
	When the position information of the active eavesdropper is completely absent at UAVs, multi-UAV applying cooperative jamming to the orthogonal space of the legitimate receiver's channel and position adjustment are effective methods to degrade the received signals at eavesdroppers and enhance that at the legitimate receiver \cite{UAV_Jamming_TVT_2018}.
	Moreover, multi-antenna techniques can be adopted to utilize spatial degrees of freedom, hence enhance the quality of the received signal at the legitimate receiver, and reduce the potential information leakage to active eavesdroppers via precise beamformer design. 
	Furthermore, one can exploit the huge bandwidth of mmWave frequency band to avoid potential eavesdropping and take advantages of line-of-sight dominated channels in mmWave systems for realizing highly directional transmission. 
	
\end{itemize}

\emph{Remark:} We note that the the aforementioned anti-jamming techniques can be also adopted to reduce information
	leakage against passive eavesdroppings. Besides, the techniques used for addressing security issues in air-to-ground channels can
	be also applied to ground-to-air ones where the UAV is a legitimate receiver. For instance, the
	flexibility of the UAV can be leveraged against eavesdropping and jamming attacks via proper
	trajectory planning or cooperative receiving.


\section{Advanced Approaches for UAV Systems}

To further enhance the system secrecy performance, some advanced techniques can also be incorporated into the UAV systems \cite{Jonas_CM_2015,5G_Book}.
In the sequel,  we will discuss the potential applications of NOMA, beamforming, and mmWave techniques in UAV systems for improving the performance of physical layer security.


\subsection{Enhancing Physical Layer Security by NOMA}

\begin{figure}[!t]
	\begin{minipage}{0.5\columnwidth}
		\centering
		\includegraphics[width=1\linewidth]{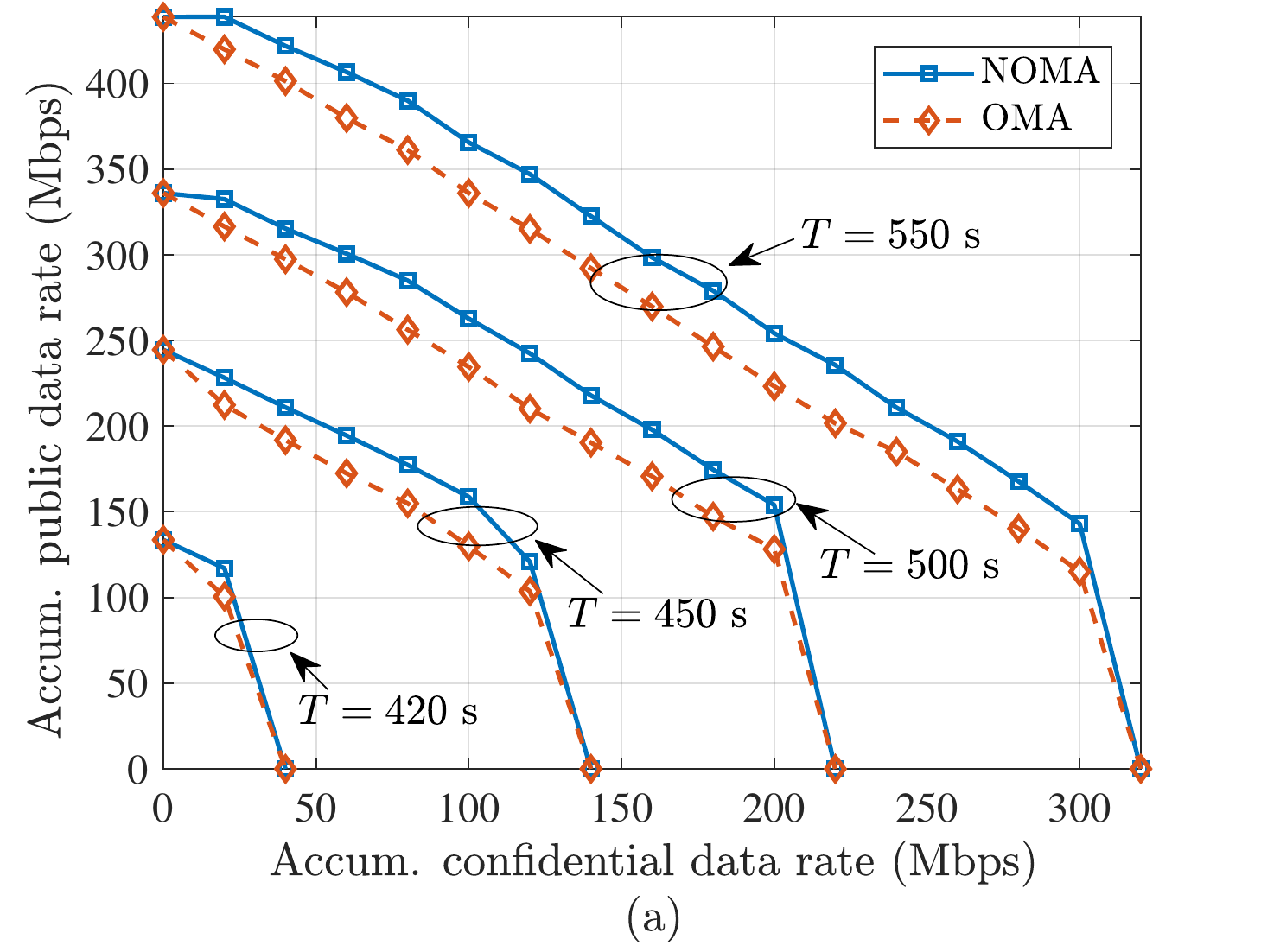}
	\end{minipage}
\begin{minipage}{0.5\columnwidth}
	\centering
	\includegraphics[width=1\linewidth]{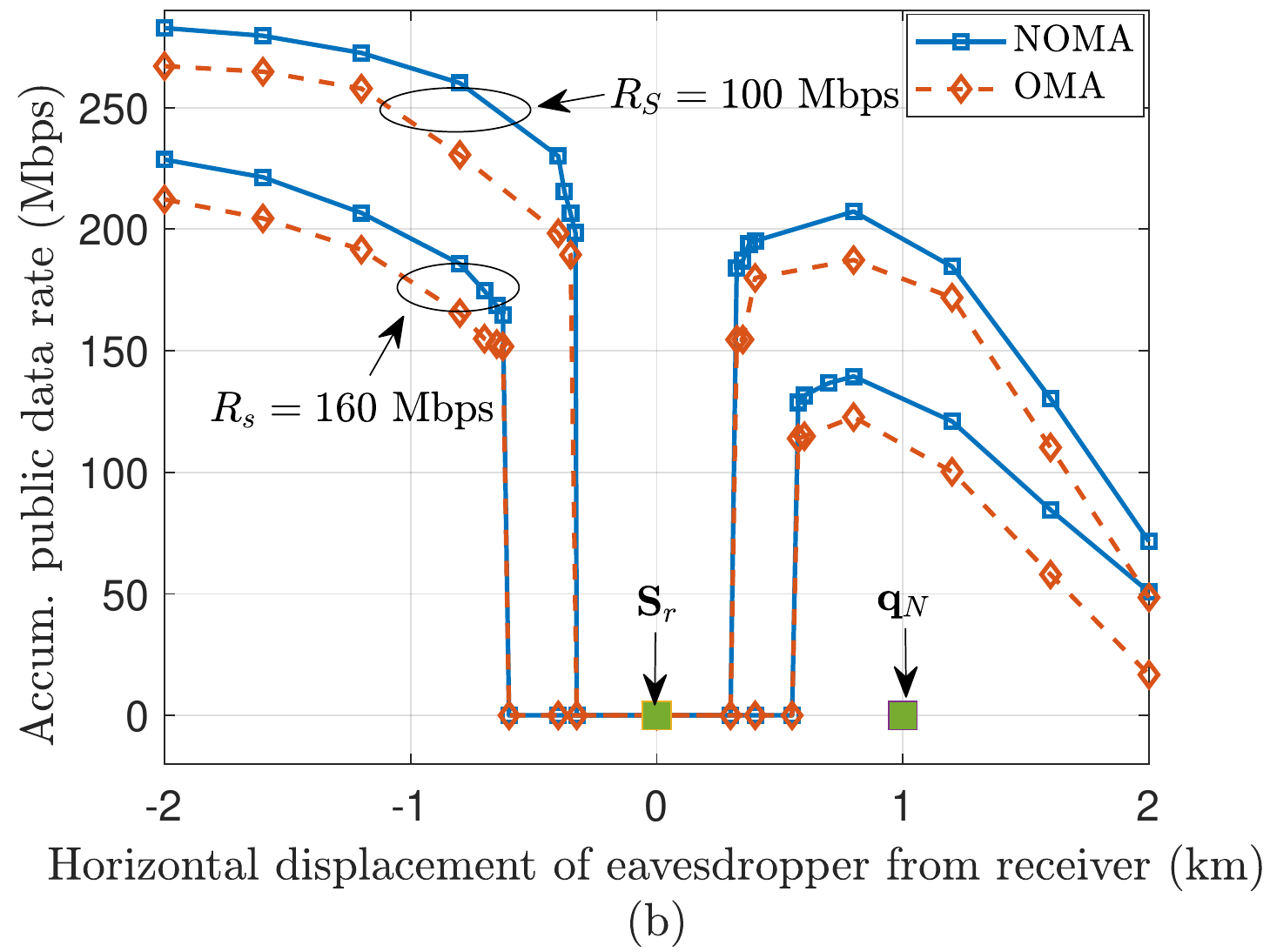}
\end{minipage}	
	\caption{(a): The rate region achieved by NOMA and OMA with different maximum cruising durations. (b): The accumulated data for the internal eavesdropper versus its displacement with different accumulated confidential data rate targets.}
	\label{fig:rate_eve_position}
\end{figure}

NOMA is viewed as a promising technique to provide superior spectral efficiency by multiplexing information signals at different power levels \cite{5G_Book}.
Hence, it is expected that NOMA can bring additional rate and robustness to enhance the achievable rate in UAV physical layer security communications.
Consider a scenario where a UAV acts as a relay to facilitate  data delivery to two receivers with different security clearance levels within a maximum cruising duration $T$.
	The receiver with a lower security clearance level is a potential eavesdropper since it has a strong motivation in intercepting signals intended to a receiver with a higher security clearance.
Then, when the eavesdropper suffers from a bad channel condition, NOMA is adopted to forward both confidential and public information simultaneously.
Otherwise, UAV only broadcasts the public information for security issues.
The mode selection between NOMA and unicast is chosen based on the results of the proposed resource allocation optimization. In particular, 
for maximizing the spectral efficiency, one needs to jointly optimize the transmission scheme, resource allocation, and UAV's trajectory.
However, the coupled optimization variables generally result in non-convex optimization problems which are difficult to solve optimally.
As an alternative, an iterative suboptimal algorithm based on successive convex approximation can be employed to facilitate a computationally efficient joint design \cite{Xiaofang_GC}.

Under the aforementioned optimization framework,
Figure \ref{fig:rate_eve_position}(a) depicts the optimal rate region achieved by NOMA and OMA, respectively, with different cruising durations.
Figure \ref{fig:rate_eve_position}(b) depicts the maximum accumulated public data rate achieved at the eavesdropper versus the horizontal displacement of the eavesdropper from the legitimate receiver.
From Figure \ref{fig:rate_eve_position}, we find that NOMA scheme always outperforms OMA in all the considered scenarios, which demonstrates the spectral efficiency advantage brought by NOMA in striking a balance between public data rate and confidential data rate.
Moreover, based on Figure \ref{fig:rate_eve_position}(b), we find that the accumulated rate drops rapidly when the eavesdropper is sufficiently close to the legitimate receiver. In fact, when the secrecy rate requirement cannot be satisfied, we set the accumulated public data rate as zero to account for the penalty of the failure in guaranteeing secure communications.

\subsection{Enhancing Physical Layer Security by  Multi-antenna Technology}

\begin{figure}[!t]
	\begin{center}
		\includegraphics[width=0.8\columnwidth]{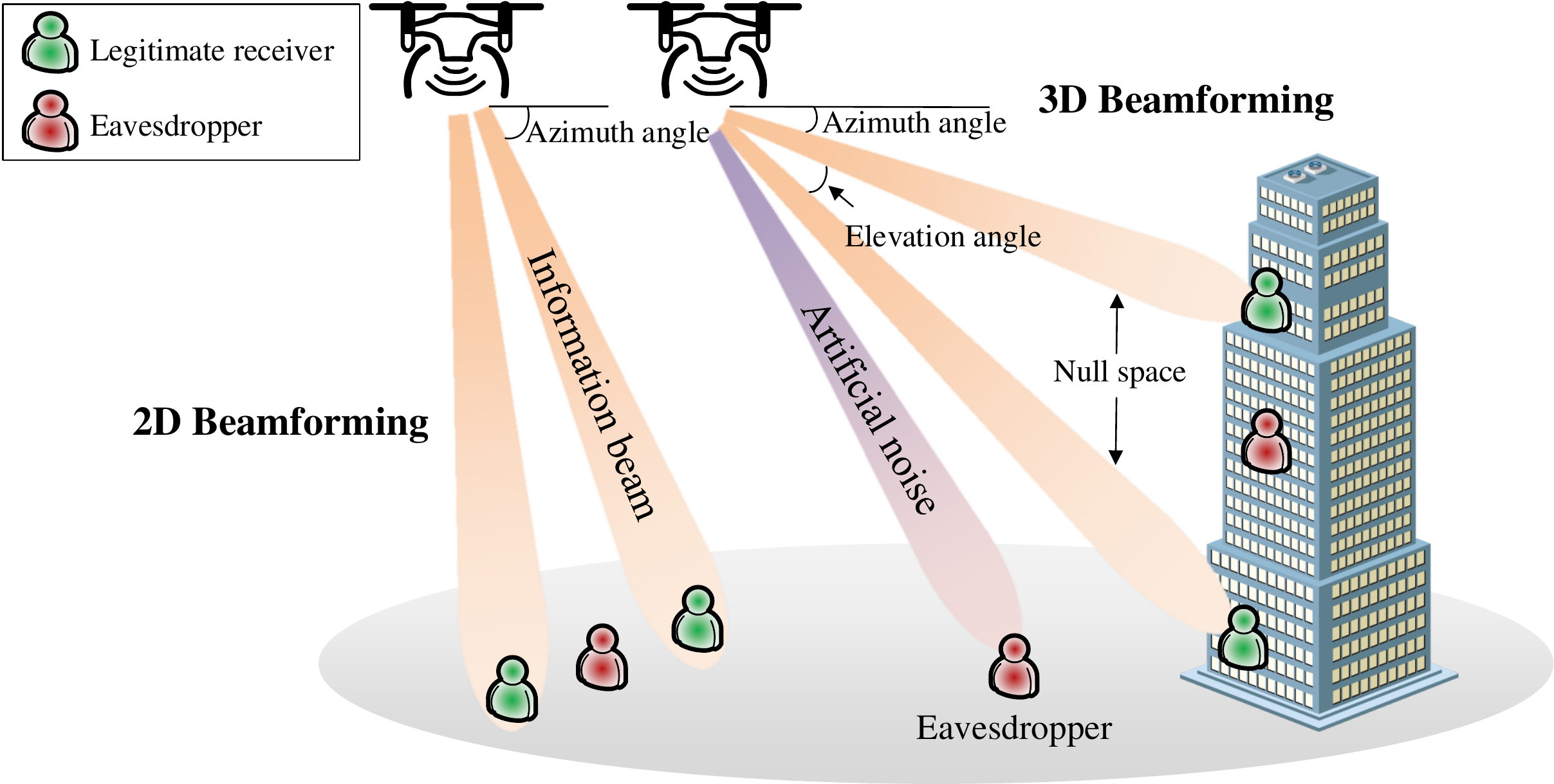}
		\caption{Illustration of the multi-antenna beamforming technique for enhancing physical layer security.}\label{fig:beamforming}
	\end{center}
\end{figure}

Multi-antenna technologies have been widely considered in wireless communications due to spatial degrees of freedom for achieving high spectral efficiency \cite{5G_Book}.
Recently, in order to concurrently improve the received signal power at the legitimate receiver and degrade the signal strength received at eavesdroppers, multi-antenna technologies have also been investigated from the physical layer security aspect to enhance the secrecy performance and robustness of the system.
In speak of the UAV wireless communication systems, the applications of multi-antenna technology can be realized from the following two approaches, cf. Figure \ref{fig:beamforming}.

\subsubsection{Traditional Two-Dimensional (2D) Beamforming}
The traditional 2D beamforming is an effective way to improve the physical layer security.
For example, when the UAV has the complete knowledge of the CSI, the beamforming direction can be simply set pointing towards the orthogonal space of eavesdropper's channel and hence the information can be safely conveyed.
However, it is obvious that this approach may not achieve the best secrecy performance of the system as the degrees of freedom are not fully utilized to improve the received signal powers at the legitimate users.
Therefore, there exists a trade-off between improving the received signal strength of the legitimate users and degrading the signal qualities of the eavesdroppers.


\subsubsection{Three-Dimensional (3D) Beamforming}
Compared to the traditional 2D beamforming, 3D beamforming can generate separated beams in the 3D space simultaneously to provide a better service coverage \cite{CellularUAV}. Thus the 3D beamforming yields a higher system throughput and can support more legitimate users than that of the 2D beamforming.
Generally, the 3D beamforming technique is more suitable for scenarios where the users are distributed in 3D space with different elevation angles to their transmitter.
Due to the high altitude of UAVs, the legitimate receivers and the potential eavesdroppers can be easily separated by their different altitudes and elevation angles to the UAVs.
Furthermore, LoS channel characteristic in UAV systems enables effective beamforming in both azimuth and elevation domains.
Specifically, narrow and precise beamformer can be created to improve the transmission efficiency with respect to the desired legitimate receivers while reduces the possibilities of information leakage to the potential eavesdroppers.

\subsection{Enhancing Physical Layer Security by mmWave}

mmWave communications have been widely studied in the literature \cite{5G_Book},  since they are able to support high data rate by utilizing the abundant frequency bands. One of the major challenges for mmWave communications is that its performance depends on the availability of LoS channels.
Thus, compared to the conventional terrestrial communications, the inherent LoS channels in UAV systems facilitate the use of mmWave for high-data rate communications.

As demonstrated in \cite{5G_Book}, the mmWave channels in UAV systems are very sparse in the angular domain, as there are generally not many scatters around UAVs in the sky.
Hence, one can take advantages of the specific channel characteristics in mmWave UAV systems for realizing highly directional transmission and exploit the huge bandwidth of mmWave frequency spectrum to avoid potential eavesdropping. 
For instance, active eavesdropping attacks in UAV systems can be addressed by adopting frequency hopping (FH) technique to hide confidential data over an ultra-wide mmWave frequency spectrum. 
	There are two main advantages of this technique.
	First, confidential data can be hided from the eavesdropper by frequently hopping to different carrier frequencies without exploiting the CSI.
	Second, frequency diversity can be exploited in FH to improve the robustness against jamming active attacks.

\section{Open Issues and Challenges}
\begin{figure}[!t]
	\begin{center}
		\includegraphics[width=1\columnwidth]{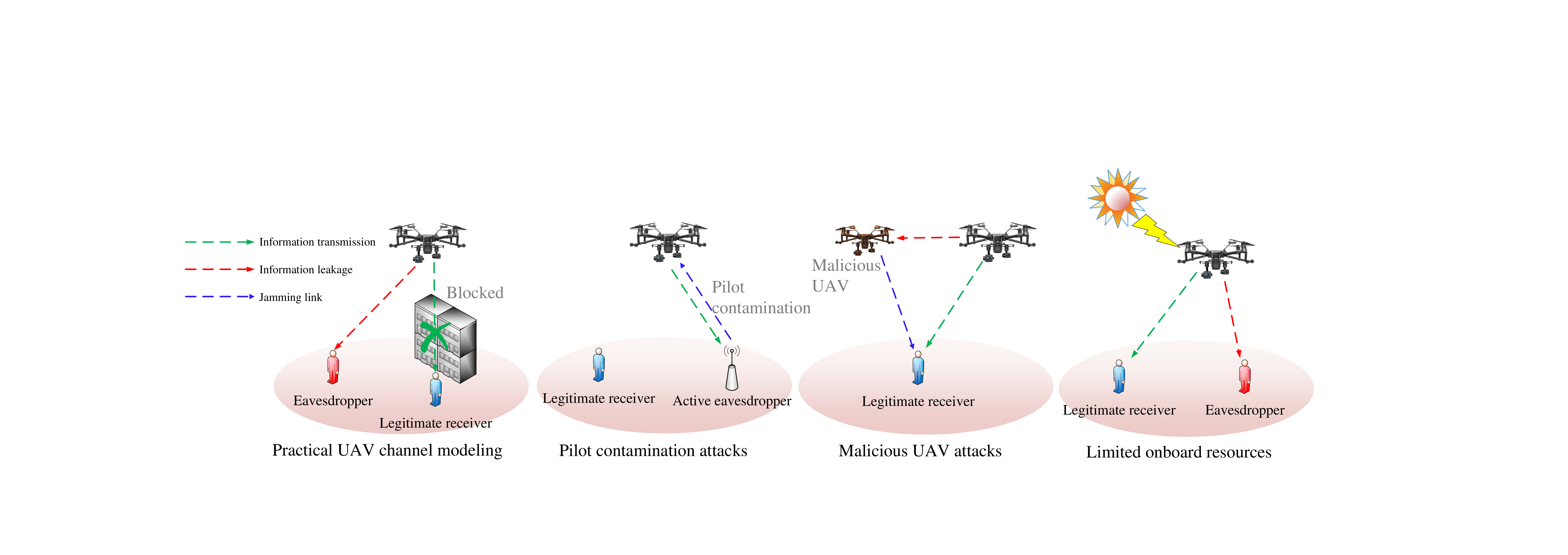}
		\caption{Illustration of security challenges in UAV wireless communications.}\label{fig:challenges}
	\end{center}
\end{figure}

UAV-based wireless communications is a growing research area and handling the associated security issues is the key to unlock its potential.
Despite the fruitful research in this area, there are variety of challenges  to be tackled, as shown in Figure \ref{fig:challenges}.
In this section, we provide and list several open issues and challenges for future works.

\subsection{Practical UAV Channel Modeling and Position Acquisition}
In the literature, air-to-ground channel is generally modeled by the LoS channel.
However, this model may not be accurate for some scenarios, such as urban areas.
Consequently, more efforts should be devoted to realistic channel modelings and to be verified by practice field test measurements.
Moreover, the mobility of UAV has significant impacts on the CSI acquisition for both the legitimate users and eavesdroppers.
This brings challenges from the physical layer security perspective, as efficient CSI acquisition algorithms usually exploit the physical properties of the wireless channels.
Therefore, it is necessary to design a pragmatic CSI acquisition mechanism to detect and track positions of the legitimate receivers and eavesdroppers.

\subsection{Physical Layer Security Against Pilot Contamination Attacks}

In practice, efficient beamforming for secure UAV communications requires accurate CSI which can be obtained by exploiting the pilot signals.
However, in some scenario, active eavesdroppers intentionally transmit deterministic pilot samples which are identical to that transmitted by the legitimate transmitter to deceive the UAV for facilitating eavesdroppings \cite{Xiangyun_Active_Pilot}.
As a result, the UAV designs an inappropriate transmission strategy which benefits the signal reception at the eavesdroppers. 
For example, the UAV may misjudge the actual network environment and fly close to the eavesdropper for confidential data delivery which increases the possibility of information leakage.
Hence, investigating some efficient policies to eliminate the impact of pilot contamination attack is challenging but important for safeguarding UAV systems.

\subsection{Physical Layer Security Against Malicious UAV Attacks}
The high mobility and flexibility of UAVs can be exploited not only to enhance physical layer security, but also to intercept the confidential data and even perform jamming to reduce the quality of the legitimate link,
which may bring more challenges than handling conventional terrestrial eavesdroppers for safeguarding UAV systems.
However, only limited researches have been devoted in this important aspect from communication theory perspective.
Therefore, it is desired to investigate advanced techniques in terms of the physical layer security to protect against malicious UAVs.

\subsection{Optimal Joint Trajectory and Resource Allocation Design}

Compare to terrestrial systems, the UAV-based system brings a new dimension to enhance physical layer security via designing the trajectory of a UAV. 
However, the designs of trajectory and resource allocation are generally coupled together, which would make the design optimization problem intractable.
Consequently, most of existing designs are suboptimal.
Unfortunately, the performance gaps between the optimal solution and existing suboptimal solutions are unclear.
Therefore, to improve the performance of physical layer security in UAV systems for mission critical applications,  efficient algorithms are desired to strike a balance between computational complexity and system performance.

\subsection{Limited Onboard Resources}

One critical obstacle in UAV wireless communication systems arises from the restricted flight duration, limited onboard energy and computational capability, etc,  as the battery capacities, sizes, and weights of UAVs are all limited.
As such, investigating advanced techniques to enable sustainable secure UAV communication are desired.
For example, energy harvesting from solar and laser provides a viable solution to supply energy to UAVs on-the-fly. Besides, UAV cooperation can effectively leverage the onboard resources among all the cooperative UAVs to enable secure communication.

\section{Conclusion}

This article provides an overview and comprehensive discussions on the specific security issues in UAV systems. 
Challenges and opportunities brought by UAVs for safeguarding UAV-based communication systems via physical layer security are fully exploited.
First, key problems in terms of security attacks are revealed. 
Then, physical layer security approaches in UAV systems are proposed to effectively prevent from both passive and active eavesdroppings.
Particularly, we provide an illustrative example on guaranteeing security provisioning by jointly designing UAV's trajectory and resource allocation, which demonstrates the advantages brought by the flexibility of UAVs.
Moreover, we propose to apply NOMA, MIMO, and mmWave techniques in UAV systems to further enhance the physical layer security and the spectral efficiency.
Finally, some potential research directions and challenges are envisioned.

\section{Acknowledgment}

This work was supported in part by funding from the UNSW Digital Grid Futures Institute, UNSW, Sydney, under a cross-disciplinary fund scheme, 
in part by the Australian Research Council's Discovery Project (DP190101363),
in part by the UK EPSRC under grant number EP/P009719/2,
and in part by H2020-MSCA-RISE-2015 under grant number 690750.

\renewcommand\refname{References}
\bibliographystyle{IEEEtran}
{\footnotesize\bibliography{IEEEabrv,Reference}}

\end{document}